\documentstyle[12pt,epsf]{article}
\begin{document}
\renewcommand{\thefootnote}{\fnsymbol{footnote}}
\begin{flushright}
KEK-preprint-95-139\\
\end{flushright}
\vskip -3cm
\epsfysize3cm
\epsfbox{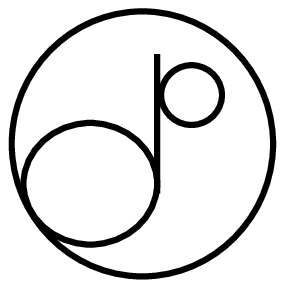}
\begin{center}
{\bf \large
Recent results from TRISTAN\footnote{
Talk presented at SLAC Summer Institute, Topical Conf. 1995.
}}\\
\vskip 0.5cm
{Ryoji Enomoto\footnote{
Internet address: enomoto@bsun03.kek.jp.}\\
}
\vskip 0.5cm
{\it
National Laboratory for High Energy Physics, KEK \\
1-1 Oho, Tsukuba-city, Ibaraki 305, Japan. \\
\vskip 0.5cm
Representing the TRISTAN Experiments\\
}
\end{center}

\begin{abstract}

The TRISTAN results from 1994 to 1995 are reviewed in this report.
The physics results dominated the $\gamma \gamma$ physics.
Therefore, only these are selected in this article.
We have systematically investigated jet productions,
the $\gamma$-structure function, and charm pair productions in
$\gamma \gamma$ processes.
The results, discussions, and future prospects are presented.

\end{abstract}

\section{TRISTAN}

\begin{figure}
\vskip -1cm
\epsfysize9cm
\epsfbox{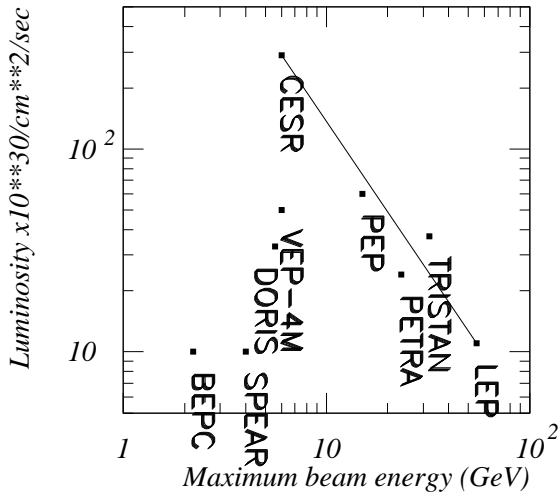}
\vskip -1cm
\caption{Luminosities versus the maximum beam energies of various $e^+e^-$
colliders.}
\label{luminosity}
\end{figure}

Initially, the TRISTAN project was
aimed at finding the ``top" quark \cite{tristan}.
Although only a three-km circumstance was available, we achieved
a maximum beam energy of 33 GeV.
Unfortunately, the top mass was far beyond this energy \cite{fnal}.
We, thus, converted our target to a high-luminosity operation of
this collider.
Figure \ref{luminosity} shows the relationship between the beam energies
and luminosities for various accelerators.
Assuming that CLEO and LEP are standard, we can see why TRISTAN
is a high-luminosity machine.
We hope that the same thing occurs in the near-future B-factory \cite{belle}.
As a matter of fact, we (TOPAZ, VENUS, and AMY) obtained
$\int L dt$ of 300pb$^{-1}$ per each experiment at $\sqrt{s}$=58 GeV.

If there is a process having a cross section
that is an increasing function of
$\sqrt{s}$, that may be a big target until the B-factory starts.
$\gamma \gamma$ physics is one of them.
The luminosity function ($L_{\gamma \gamma}$) is roughly proportional
to log($s$). As a result, TRISTAN becomes the highest luminosity
$\gamma \gamma$-factory, except for the low $W_{\gamma \gamma}$ region,
where CESR still gives the highest $\gamma \gamma$ yield.
Therefore, CESR has been fitted for resonance physics, and TRISTAN
is for parton physics. For a higher $W_{\gamma \gamma}$ region
($>6$ GeV), we have obtained the
largest statistics; this situation will remain forever.
TRISTAN can play an important role in particle physics, especially
regarding strong interactions.

\section{$\gamma \gamma$ Physics}

\begin{figure}
\epsfysize7cm
\epsfbox{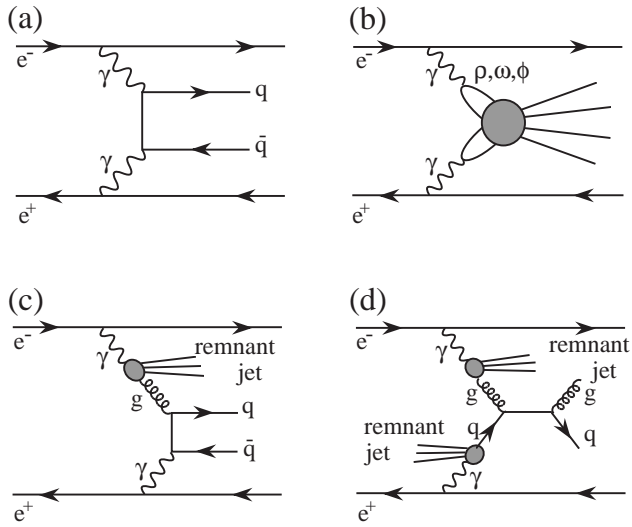}
\caption{Feynman diagrams which contributed to $\gamma \gamma$ processes
at TRISTAN.}
\label{feynman}
\end{figure}

Here, we briefly mention the $\gamma \gamma$ processes.
Figure \ref{feynman} shows four typical diagrams which contribute to these.
(a) is called a ``direct process", where photons interact with quarks via
point-like interactions \cite{direct}.
The vector-meson dominance process (VDM) is shown in (b) \cite{vdm}.
(c) and  (d) are called the ``resolved-photon process",
 where partons inside
photons interact point-like \cite{resolved}.
(a) contributes to high-$P_T$ production of quarks, (b) to a low $P_T$,
and (c), (d) to a medium $P_T$.
Considering our sensitivity over the $W_{\gamma \gamma}$ range,
in addition to our trigger system ability \cite{trigger},
we can study (a), (c), and (d) at most accurate levels.

To conclude, we are sensitive for $\gamma$-structure studies,
especially concerning the partonic structure of the photon,
in addition to higher order of QCD (or strong interaction).
The most important topic is to determine the
gluonic densities inside photons.
This is the cleanest way to determine the
$\gamma$-structure in contrast with
 ep collisions at HERA experiments.

\section{Detector}

Three groups (TOPAZ, VENUS, and AMY) were operating at TRISTAN.
Among them, we pay special attention to the TOPAZ experiments,
because of having low-angle calorimeters.

\begin{figure}
\epsfysize8cm
\vskip -1cm
\epsfbox{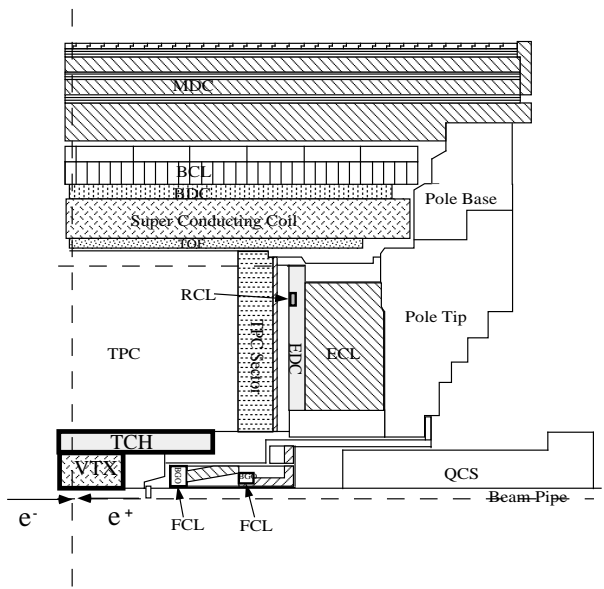}
\caption{TOPAZ detector}
\label{topaz}
\end{figure}

The apparatus of the TOPAZ detector is shown in Figure \ref{topaz}
\cite{topazref}.
The central tracker is a TPC, which enables us to study heavy flavor
productions. TOPAZ is the only detector having low-angle
calorimeters (FCL)\cite{fcl}.
This covers a polar angle region from 3.2 to 12 degrees with respect to
the beam axis.
The mean beam energy($E_b$) of TRISTAN was 29 GeV. When we select
events with an energy deposit of 0.4$E_b$ (beam-electron tag),
the $Q^2$ for the photon
is greater than 1.05 GeV$^2$.

\begin{figure}
\epsfysize7cm
\epsfbox{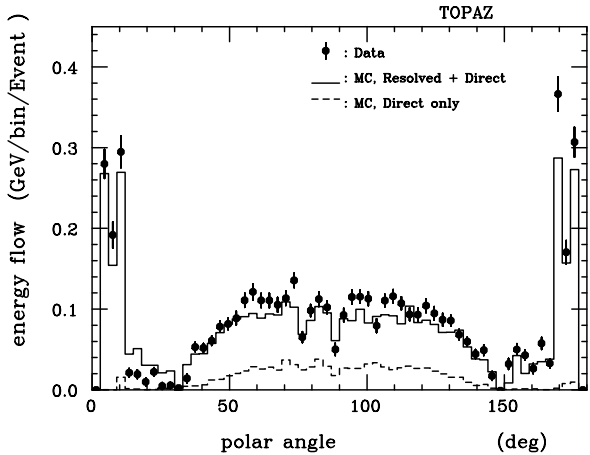}
\caption{Energy flow of $\gamma \gamma \rightarrow 2jets$ events.
The histograms are the Monte-Carlo prediction; the dashed one is the
direct process and the solid one is the resolved and direct process.}
\label{remjet}
\end{figure}

In addition to the beam-electron tag, we have introduced
a ``remnant-jet-tag". As shown in Figures \ref{feynman} (c) and (d),
hadron jets which are resolved form photons flow into beam
directions. Typically, hadrons from these jets have $P_T$'s
of about 0.4 GeV. Assuming that
these hadrons have energies of several GeV,
they hit the FCL fiducial region. The energy flow in
typical $\gamma \gamma \rightarrow 2jets$ events are shown in
Figure \ref{remjet}.
It has enhancements at low-angle regions which cannot be explained
by the processes shown in Figures \ref{feynman} (a) nor (b).

\begin{figure}
\vskip -1cm
\epsfysize9cm
\epsfbox{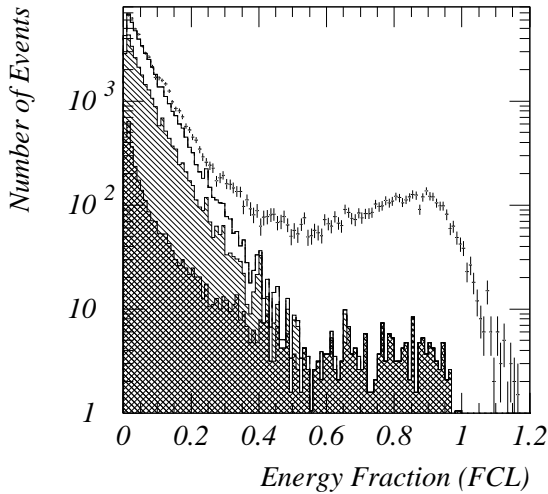}
\vskip -1cm
\caption{
Distribution of the energy fractions (normalized at the beam energy)
of the maximum-energy clusters in FCL.
The points with error bars are experimental data.
The histograms are predictions by a Monte-Carlo simulation;
the cross-hatched area is a single-photon-exchange process,
the singly-hatched one is VDM, and the open one is a resolved-photon
process.}
\label{fclhad}
\end{figure}

The energy deposits in the FCL are also shown in Figure \ref{fclhad}.
The soft component corresponds to these resolved-photon events.
We can, therefore, tag the resolved-photon process by selecting
a soft energy deposit in the FCL.
The efficiency of this tagging was estimated to be $\sim$80\%
with a background of 10\%, mostly due to the beam background.
We call this ``remnant-jet-tag", or ``rem-tag" in short.

\section{Event Structure}

\subsection{Event Shapes}

\begin{figure}
\epsfysize7cm
\epsfbox{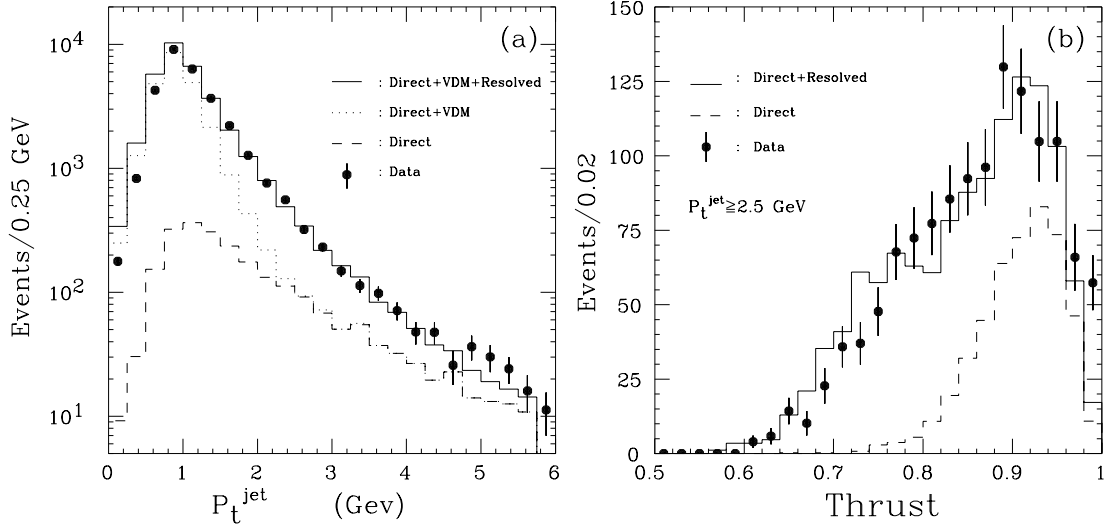}
\caption{(a) $P_T^{jet}$ distribution. The histograms are theoretical
predictions; the dashed one is a direct process, the dotted one is the direct
and VDM, and the solid one is the sum of these two and the resolved-photon
process. (b) Thrust distribution of high $P_T^{jet}$ ($>2.5$ GeV) events.}
\label{evshape}
\end{figure}

As has been described, various processes contribute to $\gamma \gamma$
collisions; the analysis ways are not unique.
According to a historical method, hadron system at the CMS frame were
divided into two hemispheres (definition of jets).
This method has an advantage for analyzing events in all $P_T$ regions.
AMY showed evidence for a resolved-photon process by this method
\cite{tanaka}.
A similar analysis was carried out by TOPAZ, and the $P_T^{jet}$
distribution is shown in Figure \ref{evshape} (a).
For example, at $P_T^{jet}$=2.5 GeV, the data excess is by a factor of 2.5
compared with the incoherent sum of direct and VDM processes.
This excess has been explained by the resolved-photon process.
Next, the thrust distribution of high $P_T^{jet}$ ($>$2.5 GeV)
events are plotted in Figure \ref{evshape} (b). The events
are spherical, consistent with the prediction by the resolved-photon
processes. Similar results have been obtained by the LEP experiments
\cite{aleph,delphi}.

\subsection{Jet Cross Section}

\begin{figure}
\epsfysize7cm
\epsfbox{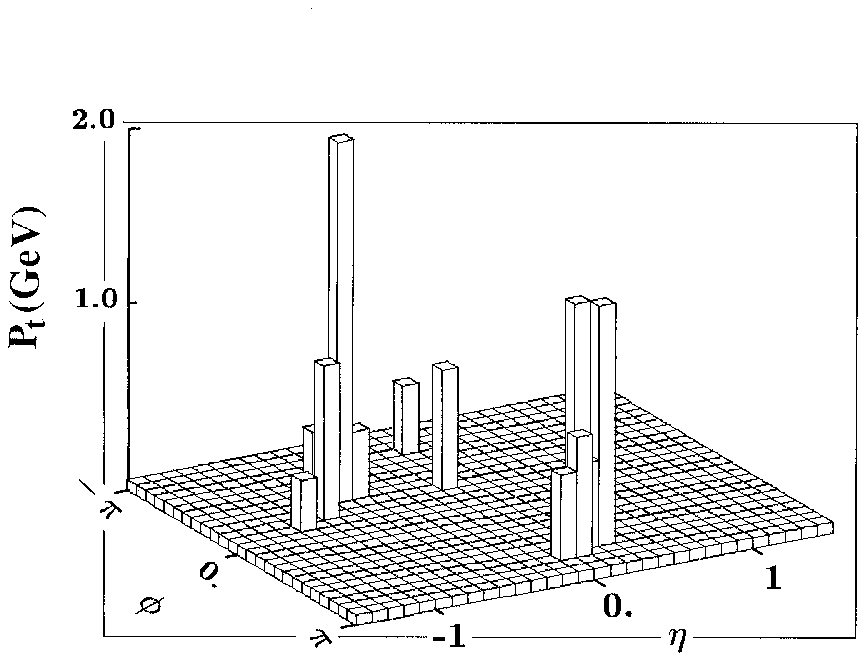}
\caption{Typical jets observed by the TOPAZ detector.}
\label{jet}
\end{figure}

\begin{figure}
\epsfysize7cm
\epsfbox{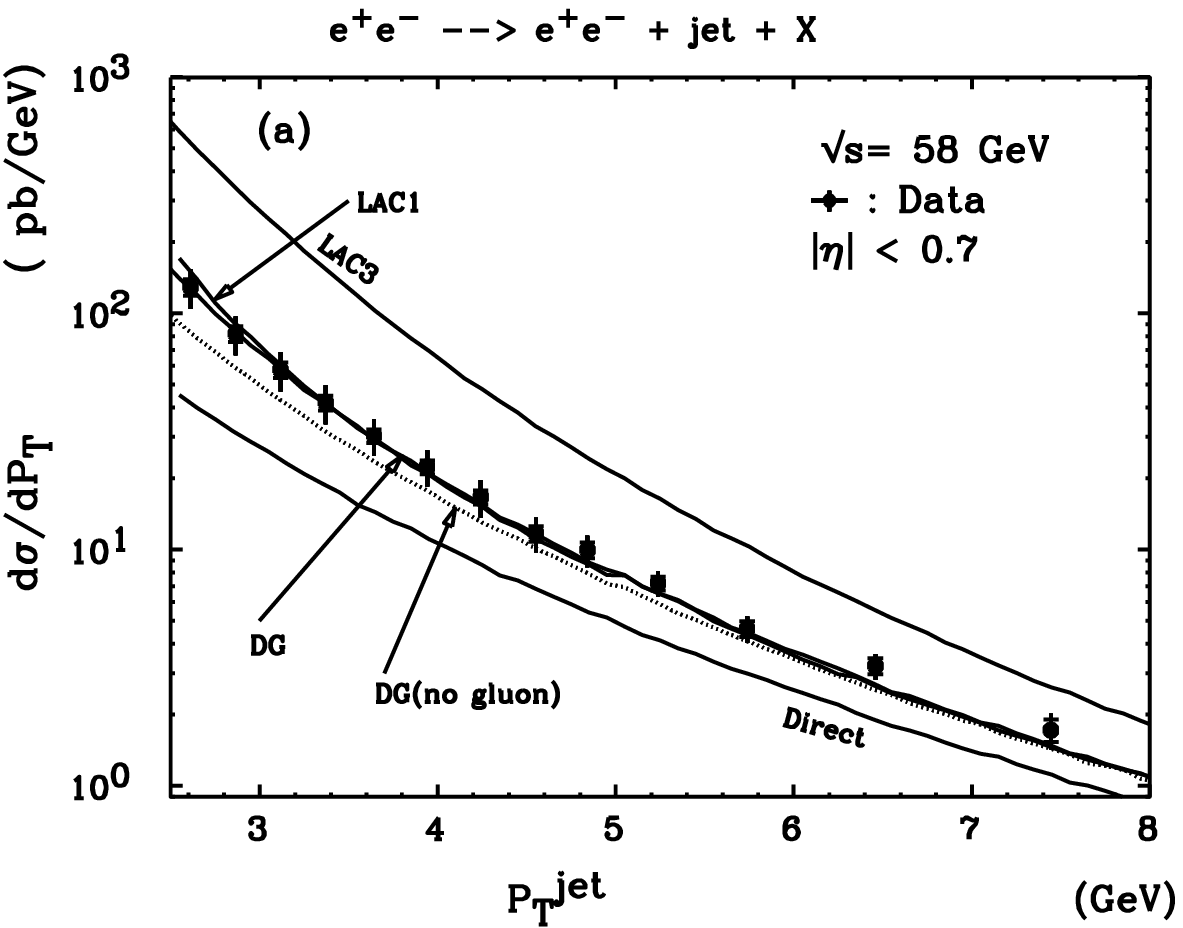}
\caption{Jet production cross section in $\gamma \gamma$ collisions.}
\label{jetcross}
\end{figure}

The processes shown in Figures \ref{feynman} (a), (c), and (d)
include hard scattering of partons which are observed as jets
(Figure \ref{jet}).
These jets are reconstructed in $\phi$ and $\eta$ plane. The
particles within the circle $R=\sqrt{(\phi-\phi_0)^2
+(\eta-\eta_0)^2}$ are used.
Figure \ref{jetcross} is the cross section of jet production versus
$P_T^{jet}$.
The cross section is consistent with the incoherent sum of the
direct and resolved-photon processes at the $P_T^{jet}>$2GeV
region (the same result as the previous one).
The theoretical models, called LAC1, LAC3, and DG, shown in the
Figure have significant differences in the gluon distribution inside
the photon \cite{lac,dg}
The hard-gluon model (LAC3) is clearly rejected.
LAC1 and DG show difference at low-x gluonic-density,
and it is difficult to distinguish them by this experimental
method \cite{hayashii}.
A similar result was obtained by AMY \cite{kim}.

\section{Structure Function}

The photon structure function ($F_2^{\gamma}$) was measured by the TOPAZ
collaboration \cite{muramatsu}.
We obtained a high value compared with the
theoretical values at x$\sim$0.04 at
$3<Q^2<30$ GeV$^2$.
These regions are important for determining the QCD-based models.
Although the experimental ambiguities in determining x value
from the mass of the measured hadronic system were found to be large,
and there will be a systematic shift.
We are, therefore, going to reanalyse the data using a new
algorithm to determine x while assuming missing-energy flow directions
(i.e., beam-pipe direction).

\section{Charm Pair Production}

According to a QCD calculation of parton-parton scattering,
the cut-off parameter ($P_T^{min}$) was introduced for light-quark
scattering. This parameter must be determined experimentally;
the optimum value was obtained to be around 1.7$\sim$2 GeV.
Fortunately, for the charm-quark case, this parameter is not necessary,
and we can experimentally select charm-pair events with high
purity.
In addition, the VDM effects are considered to be small for
charmed-particle production.
In the resolved-photon processes we only have to consider
gluon-gluon scattering; therefore, this is sensitive to the
gluonic density in which a large model dependence exists.
The NLO calculations are available at the parton level \cite{nlo}.

\subsection{Full and partial Reconstruction of $D^{*\pm}$}

Initial charm quark fragment to $D$ mesons. $D^*$ is the most
probable state. This fragmentation function is experimentally
well known. We can, therefore, estimate the initial charm quarks' $P_T$s.

\begin{figure}
\vskip -1cm
\epsfysize9cm
\epsfbox{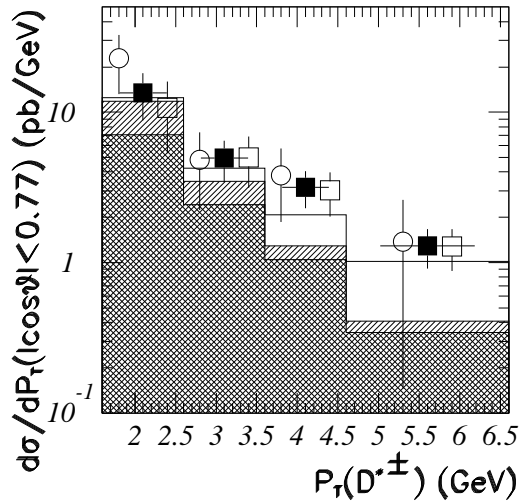}
\vskip -1cm
\caption{Differential cross sections of $D^{*\pm}$ versus $P_T$.
The open circles were obtained by the full reconstruction method,
the open squares by the softpion method, and the closed squares
are the average by these two methods.
The histograms are the theoretical predictions: the cross hatched
is the direct process, the singly hatched is the resolved, and the
open area is obtained assuming the hypothetical $\tilde{t}$ pair
production.}
\label{dstarcrs}
\end{figure}

We first tried to reconstruct $D^{*+}\rightarrow \pi_s^+D^0 (D^0\rightarrow
K^-\pi^+X)$ \cite{dstar}.
$20\pm5$ $D^*$s were reconstructed with a good S/N ratio. The obtained
cross section is plotted in Figure \ref{dstarcrs} by open circles.
Although
this cross section is higher than the sum of the direct and resolved-
photon predictions, the statistics are low.
The NLO effect was taken into account in the prediction.
We used LAC1 for the gluonic density in photons.


In order to improve the experimental accuracy, we carried out a
``softpion analysis" in reconstructing the $D^*$s.
The results are also shown in Figure \ref{dstarcrs} by the open circles.
They are consistent with that of the full reconstruction.
The high $P_T$ anomaly still existed, and the hypothetical $\tilde{t}$
pair assumption was tested by looking at the event shapes \cite{softpi}.
These shapes differ from $\tilde{t}$-pair prediction and
rather resemble the typical $\gamma \gamma$ events.
The similar high $P_T$ anomaly was also reported by the AMY
collaboration \cite{aso}.

\subsection{$K_s$ Inclusive}

The maximum integrated luminosity of the TRISTAN experiment is 300pb$^{-1}$
and now most of them were analyzed. We must, therefore, seek other ways
of analysis than waiting for an increase in data.
An inclusive analysis of the strange particle is one of them.
The $P_T$ spectrum of these reflect that of charm quarks.
Also, strange-quark pair production is strongly suppressed in
$\gamma \gamma$ collisions.

\begin{figure}
\epsfysize7cm
\epsfbox{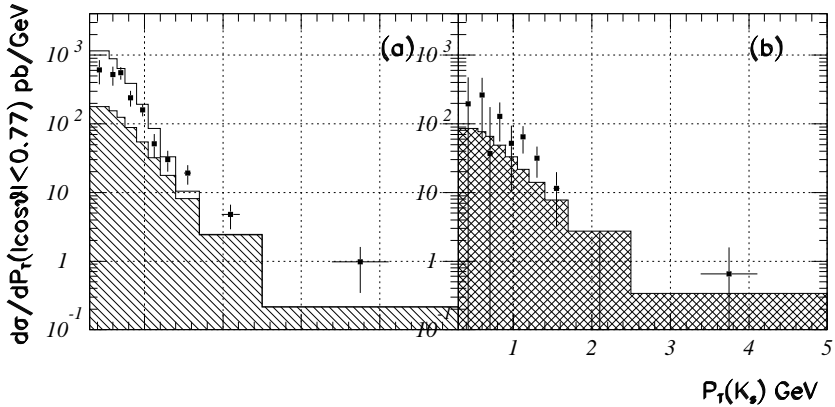}
\caption{Differential cross sections of $K_0$ versus $P_T$:
(a) for the VDM and resolved-photon process, and (b) for the direct process.
The histograms are the theoretical predictions; the definition of the hatches
are as same as those for the previous figure.}
\label{kscrs}
\end{figure}

In the $K_s$ inclusive analysis, we introduced ``remnant-jet-tag" \cite{ks}.
The details were described in the previous section.
We can, therefore, derive the cross sections process by process.
These are shown in Figures \ref{kscrs} (a) and (b).
Here, we could not separate the VDM and the resolved-photon events,
because of the low-$P_T$ particle production by the VDM.

By this study, the existing theory plus the LAC1 parametrization
with the NLO correction well describe the experimental data.
We further tested the parametrization difference in the gluon density
by using the WHIT parameterization \cite{whit}.
This gave six systematic parametrizations. Some combinations
of these with various $P_T^{min}$ cut-offs, fitted the experimental data
perfectly.

\subsection{Electron Inclusive}

\begin{figure}
\epsfysize7cm
\epsfbox{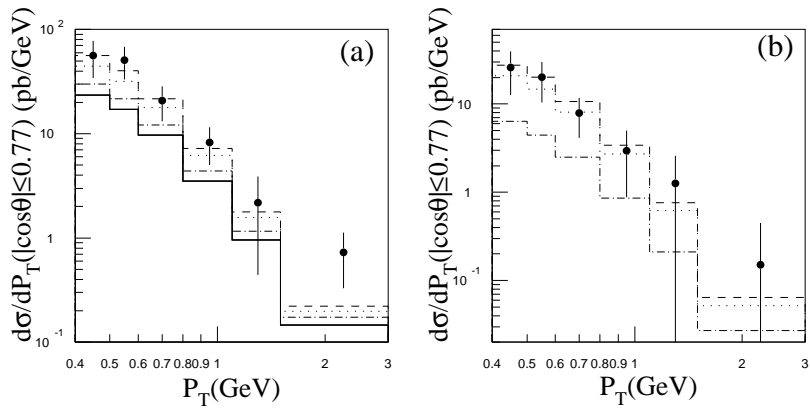}
\caption{(a) Electron inclusive cross sections; the solid line is the
direct process, the dot-dashed is the DG case, the dotted is the LAC1
with $m_c$=1.5 GeV, and the dashed is the LAC1 with $m_c$=1.3 GeV.
(b) The resolved-photon cross section; the definition of lines are as
same as (a).}
\label{elect}
\end{figure}

The electron inclusive method is a cleaner one than the inclusive $K_0$.
Here, we do not have to consider the VDM.
The TOPAZ detector can identify very low $P_T$ electrons, such as 0.4 GeV
\cite{elecnim}.
We can, therefore, measure the gluon density at very low-x ($\sim$0.02),
where the model differences appears.
Figure \ref{elect} (a) is the differential cross section versus
the electron $P_T$s.
The experimental data clearly favor the LAC1 parametrization, also
suggesting the necessity of the NLO correction and a low charm-quark mass
of $\sim$ 1.3 GeV \cite{iwasaki}. Although
VENUS produced a similar result, the statistics were about
half that of ours \cite{uehara}.

We carried out ``remnant-jet-tag", and obtained purely
``resolved-photon" cross section (Figure \ref{elect} (b)).
Again, it confirmed our parametrization of the theory.
We also observed a large difference between the DG and LAC1. This is
because this method is sensitive to very low-x regions
where the jet analysis could not resolve.
Note that
this method is more powerful than the single tag experiment ($F_2^{\gamma}$)
in determining the gluon density inside photons.

In figure \ref{elect} (a) at highest $P_T$ region, there are some
excesses compared with the existing theory.
A similar high $P_T$ excess was observed by AMY \cite{aso}.

\subsection{$\Lambda$ Inclusive}

So far what we have learned is that there are some high $P_T$ excess
in charm production, and that the experimental results at low
$P_T$ agree with the existing theory with the NLO correction and high
gluonic density at low x.
We investigated the $\Lambda$-inclusive cross section in order to
qualitatively study
the NLO effects.
$\Lambda$s can also tag charm-pair events the same as in the $K_0$ case.

In addition, there is an experimental fact that a gluon-jet produces
more $\Lambda$s than does a quark jet \cite{cleoinc}.
Our experimental results are shown in Table \ref{lamratio}
\cite{lambda}.
\begin{table}
\begin{tabular}{ccccc}
\hline
\hline
tag cond. & Experiment & Theory (LO) & Exp./Theory & subprocess \\
\hline
antitag & 43.3$\pm$8.3 & 19.1 & 2.26$\pm$0.43 & VDM+resolved+direct \\
rem-tag (-VDM) & 15.6$\pm$3.5 & 6.0 & 2.60$\pm$0.58 & resolved \\
rem-tag & 34.8$\pm$7.8 & 17.3 & 2.01$\pm$0.45 & VDM+resolved \\
anti-rem & 27.7$\pm$7.9 & 13.1 & 2.11$\pm$0.60 & VDM+direct \\
\hline
\hline
\end{tabular}
\caption{
Total cross section (pb) of $\Lambda (\overline{\Lambda})$ in the
$|\cos \theta |<0.77$ and $0.75<P_T<2.75$ GeV range.
The notation (-VDM) means VDM subtraction
using theory.
Here, we use the LO theories in order to show the discrepancy
with the experimental data.
}
\label{lamratio}
\end{table}
There are process-independent excesses compared with the prediction of
the LO theories. The values are a factor of two. We can, therefore, conclude
that there exists significant gluon jet production in $\gamma \gamma$
collisions, i.e., the NLO effect.

\section{Double Tag}

\begin{figure}
\epsfysize7cm
\epsfbox{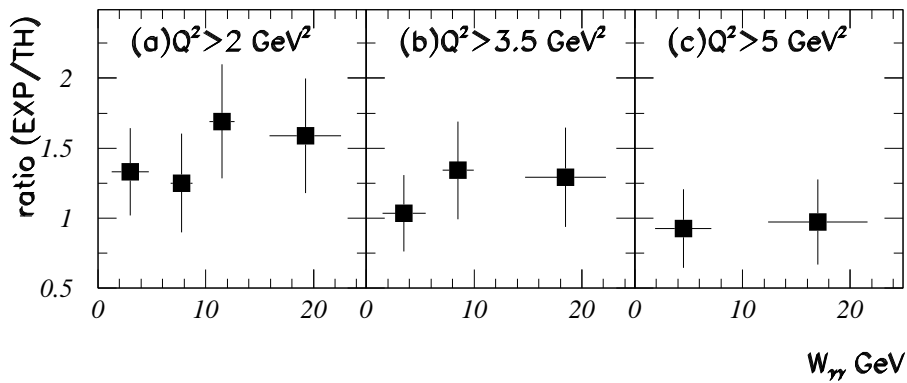}
\caption{
Ratios of the experimental and theoretical cross sections
in
various kinematic regions;
(a) $Q^2_{\gamma ,min}>$ 2GeV$^2$,
(b) $Q^2_{\gamma ,min}>$ 3.5GeV$^2$,
and (c) $Q^2_{\gamma ,min}>$ 5GeV$^2$.
The $Q^2_{\gamma , min}$ is specified in the text.
}
\label{dtratio}
\end{figure}

We carried out a double-tag analysis, and obtained the total hadronic
cross sections \cite{dtag}.
The $Q^2$ ranges for $\gamma ^*$ was 2$\sim$25 GeV$^2$ and
the $W$ range was  2$\sim$25 GeV.
Figure \ref{dtratio} is the ratio of the cross sections ($e^+e^-\rightarrow
e^+e^-h$) between the experiment and the LO $e^+e^-\rightarrow
e^+e^-q\bar{q}$ theory.
The experimental value agrees with the LO prediction in the high-$Q^2$
region. There are enhancements of $\sim$30\% in the low-$Q^2$
region, suggesting NLO effects.

\section{Discussion}

Our experimental data strongly favor a large gluon density at low x,
as has been suggested by LAC1.
However, the HERA experiment (ep collision) showed a lower gluon
density than that which LAC1 predicted \cite{hera}.
Also, the LEP experiment is inconsistent with LAC1 at low x \cite{lep}.
The problem is whether we can explain all of the experimental data by
simply changing the parton density functions.
In addition, the high $P_T$ excess in charm production can not be
solved by any existing theories.

The cross section of the $\gamma \gamma$ collision increases with energy
in a future $e^+e^-$ linear collider experiment.
It would be a large background, and may be related to such physics
as $H\rightarrow \gamma \gamma$ searches.
In order to reliably estimate the background, our measurement
greatly helps.
Systematic measurements, such as $\gamma \gamma \rightarrow h^{\pm}X$
and $\gamma X$, are necessary.

\section{Conclusion}

At the TRISTAN $e^+e^-$ collider, a systematic study of
hadronic $\gamma \gamma$ collisions was carried out.
TRISTAN is a high-luminosity $\gamma \gamma$ factory, and our
data of these processes have the largest statistics.
For parton production our data greatly contributed to our
experimental and theoretical understanding
of photon structures.
Further systematic measurements on various processes are awaited.

\section*{Acknowledgements}

I thank to Drs. H. Hayashii, M. Iwasaki, and T. Nozaki for summarizing
these results.
I also thank to the TOPAZ collaboration and the KEK accelerator
division.

\end{document}